\documentclass[aps,prd,amsfonts,preprint, eqsecnum,nofootinbib]
{revtex4} 
\usepackage{graphics,epsfig}
\begin{document}
\preprint{WM-03-105}
%
\title{\vspace*{0.5in} Phenomenology of the Pentaquark Antidecuplet
\vskip 0.1in}
\author{Carl E. Carlson} \email[]{carlson@physics.wm.edu}
\author{Christopher D. Carone}\email[]{carone@physics.wm.edu}
\author{Herry J. Kwee}\email[]{herry@camelot.physics.wm.edu}
\author{Vahagn Nazaryan}\email[]{vrnaza@wm.edu} 
\affiliation{Nuclear and Particle Theory Group, Department of Physics,
College of William and Mary, Williamsburg, VA 23187-8795}
\date{July 2003}
\begin{abstract}
We consider the mass splittings and strong decays of members of the 
lowest-lying  pentaquark multiplet, which we take to be a parity-odd 
antidecuplet.  We derive useful decompositions of the quark model wave 
functions that allow for easy computation of color-flavor-spin matrix 
elements.  We compute mass splittings within the antidecuplet including 
spin-color and spin-isospin interactions between constituents and point 
out the importance of hidden strangeness in rendering the nucleon-like states 
heavier than the S=1 state.  Using recent experimental data on a possible 
S=1 pentaquark state, we make decay predictions for other members of the 
antidecuplet. 
\end{abstract}
\pacs{}
\maketitle

\section{Introduction}\label{sec:intro}

Recently, a number of laboratories have announced observation of a
strangeness $+1$ baryon~\cite{nakano,barmin,stepanyan} with a mass of
1540 MeV and a narrow decay width.  Such a state cannot be a 3-quark
baryon made from known quarks, and it is natural to interpret it as a
pentaquark state, that is, as a state made from four quarks and one
antiquark, $q^4 \bar q$. The current example of the strangeness $S=+1$
baryon is positively charged and is called $Z^+$ in the particle data
tables and $\theta^+$ in some recent works~\cite{stepanyan}.  The $Z^+$
of necessity has an $\bar s$ and four non-strange quarks.  The parity,
spin, and isospin of the experimental state are currently unmeasured.

In this paper, we study consequences of describing the $Z^+$ within the 
context of conventional constituent quarks models, in more focused 
detail than was done in earlier work~\cite{hogaasen,strottman,roiesnel} 
and with new results.  In these models, all quarks  are in the same 
spatial wave function, and spin dependent mass splittings come from 
either color-spin or flavor-spin exchange.  The $Z^+$ made this way has 
negative  parity.  We treat it as a flavor antidecuplet, with spin-1/2 
because this state has, at least by elementary estimates, the lowest 
mass by a few hundred MeV among the $Z^+$'s that can be made with all 
quarks in the ground spatial state.

The pentaquark by now has some history of theoretical study.  In the
context of constituent quark models, it was analyzed relatively early
on~\cite{hogaasen,strottman,roiesnel}, but the subject was not
pursued, probably for lack of experimental motivation. (The first
of~\cite{hogaasen} gives a simple estimate of the $Z^+$ mass of 1615
MeV and then states ``There definitely is no $Z^*(I=0)$ state at such
a low mass.'')  Much of the effort shifted to studying pentaquarks
involving charmed as well as strange
quarks~\cite{heavyexamples,leandri}, before the recent flurry of
theoretical attention~\cite{recent}.

Pentaquarks have also been studied in the context of the Skyrme 
model~\cite{oh,diakonov}.  Ref.~\cite{diakonov} in particular makes a 
striking prediction, based on the assumption that the $Z^+$ is a member of a 
flavor antidecuplet and that the nucleon-like members of this decuplet 
are the observed $N^*(1710)$ states, that the $Z^+$ would have a mass 
of about 1530 MeV and a width less than 15 MeV.  Note that in this case 
the $Z^+$ is a positive parity state.

We may elaborate on the $Z^+$ states and masses in quark models 
briefly before proceeding. In outline, there are several ways to make a 
$Z^+$, and one can obtain $Z^+$'s which are isospin 0, 1, or 2.  The 
mass splittings between the states can be estimated using, say, the 
color-spin interactions described in more detail in the next 
section.  Techniques and useful information may be found 
in~\cite{hogaasen,leandri,jaffe}.  The lightest $Z^+$ state is the 
isosinglet (in the $\overline{\bf 10}$) with spin-1/2.  The isosinglet 
spin-3/2 is a few hundred MeV heavier.  The heaviest states are the 
isotensor spin-1/2 and (somewhat lighter) spin-3/2 states.  The mass 
gap between the lightest and heaviest of the $Z^+$'s is triple the mass 
gap between the nucleon and the $\Delta(1232)$, if one does not account 
for changes in the quarks's spatial wave functions ({\em e.g.}, due to 
changes in the Bag radius), or the better part of a GeV.  The isovector 
masses lie in between the two limits. These statements are considered
in quantitative detail in Ref.~\cite{cckv2}

In the next section, we will discuss the color-flavor-spin wave 
functions of the antidecuplet that contains the $Z^+$.  This is a 
necessary prelude to a discussion of the mass splittings and decays of 
the full decuplet, which follows in Section III.  One intriguing  
result is the roughly equal mass spacing of the antidecuplet, with the 
$Z^+$ lightest.  Normally one expects the strange state to be heavier 
that the non-strange one.  The explanation of this counterintuitive 
behavior is hidden strangeness, that is, there is a fairly high 
probability of finding an $s \bar s$ pair in the non-strange state.  
We also show that there is a markedly different pattern of
kinematically allowed decays, depending of whether spin-isospin or
spin-color exchange interactions are relevant in determining the
mass spectrum. We close in Section IV with some discussion.

\section{Wave Function}\label{sec:wf}

\begin{figure}[t]
\epsfxsize 3.5 in \epsfbox{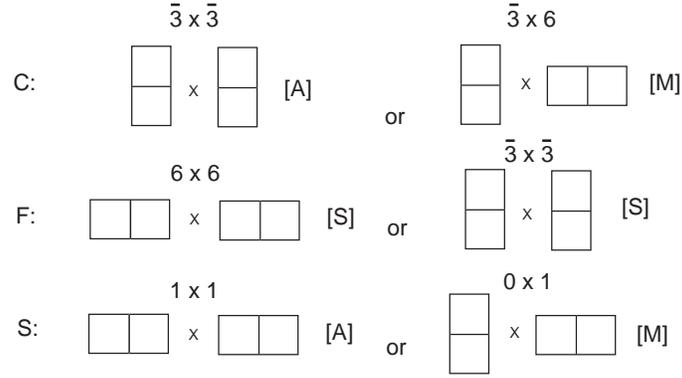} \caption{Quark pair states
that can be appropriately combined to yield a total (C,F,S) 
state $({\bf 3},{\bf \bar 6}, 1)$. }
\label{fig:yt}
\end{figure}

There are two useful ways to compose the pentaquark state.  One is to
build the $q^4$ state from two pairs of quarks and then combine with
the $\bar{q}$.  The other is to combine a $q^3$ state with a
$q\bar{q}$ to form the pentaquark. We first represent the pentaquark
state in terms of states labelled by the quantum numbers of the first
and second quark pairs.  Since the antiquark is always in a (${\bf
\bar 3}$,${\bf \bar 3}$,$1/2$) (color,flavor,spin) state, we know
immediately that the remaining four-quark ($q^4$) state must be a
color ${\bf 3}$.  The flavor of a generic $q^4$ state can be either a
${\bf 3}$, ${\bf \bar 6}$, ${\bf 15}_M$, or ${\bf 15}_S$ (where $S$
and $M$ refer to symmetry and mixed symmetry under quark interchange,
respectively).  However, only the ${\bf \bar 6}$ can combine with the
${\bf \bar 3}$ antiquark to yield an antidecuplet. Finally, the spin
of the $q^4$ state can be either $0$ or $1$ if the total spin of the
state is $1/2$.  However, it is not difficult to show that any state
constructed with the correct quantum numbers using the spin-zero $q^4$
wave function will be antisymmetric under the combined interchange of
the two quarks in the first pair with the two quarks in second pair;
this is inconsistent with the requirement that the four-quark state be
totally antisymmetric.  Thus we are led to the unique choice
\begin{equation}
|(C,F,S)\rangle_{q^4} = |({\bf 3},{\bf \bar 6}, 1) \rangle  \,\,\, .
\end{equation}
Figure~\ref{fig:yt} shows the possible quark pair combinations that
can provide a $({\bf 3},{\bf \bar 6}, 1)$ four-quark state. The
symmetry under interchange of quarks $1$ and $2$, or $3$ and $4$ is
immediate from each of the Young's Tableau shown.  The symmetry under
interchange of the first and second quark pairs is indicated in
brackets next to the tableau.  Only three combinations have the right
symmetry under quark interchange to form a totally antisymmetric $q^4$
state, namely
\[
|({\bf \bar 3},{\bf 6}, 1)({\bf \bar 3},{\bf 6}, 1) \rangle
\,\,\, , \,\,\, \frac{1}{\sqrt{2}}\left(|({\bf 6},{\bf 6}, 0)({\bf
\bar 3},{\bf 6}, 1) \rangle + |({\bf \bar 3},{\bf 6}, 1) ({\bf 6},{\bf
6}, 0) \rangle \right) \,\, , \,\,
\]\[
\frac{1}{\sqrt{2}}\left(|({\bf 6},{\bf \bar 3}, 1)({\bf \bar
  3},{\bf \bar 3}, 0) \rangle + |({\bf \bar 3},{\bf \bar 3}, 0)({\bf
  6},{\bf \bar 3}, 1) \rangle\right) \,\,\, .
\]
The requirement of total antisymmetry of the $q^4$ wave function,
determines the relative coefficients.  We find that the properly
normalized state is given by
\begin{eqnarray}
| ({\bf 1},{\bf \overline{10}}, 1/2) \rangle &=& \frac{1}{\sqrt{3}} 
|({\bf \bar 3},{\bf 6}, 1)({\bf \bar 3},{\bf 6}, 1) \rangle
+\frac{1}{\sqrt{12}} \left(|({\bf 6},{\bf 6}, 0)({\bf
\bar 3},{\bf 6}, 1) \rangle + |({\bf \bar 3},{\bf 6}, 1) ({\bf 6},{\bf
6}, 0) \rangle \right) \nonumber \\
&-&\frac{1}{2} \left(|({\bf 6},{\bf \bar 3}, 1)({\bf \bar
  3},{\bf \bar 3}, 0) \rangle + |({\bf \bar 3},{\bf \bar 3}, 0)({\bf
  6},{\bf \bar 3}, 1) \rangle\right) \,\,\, ,
\label{eq:221}
\end{eqnarray}
where we have suppressed the quantum numbers of the antiquark, (${\bf
\bar 3}$,${\bf \bar 3}$,$1/2$), which are the same in each term.  Also
tacit on the right-hand side is that each $q^4$ state is combined to
$({\bf 3},{\bf\bar{6}},1)$.  The signs shown in Eq.~(\ref{eq:221})
depend on sign conventions for the states on the right-hand side.  For
the $Z^+$ component,spin $\uparrow$, we find
\begin{eqnarray}
|({\bf \bar 3},{\bf 6}, 1)({\bf \bar 3},{\bf 6}, 1) \rangle  &=& 
\frac{1}{24 \sqrt{3}} (c_1^j c_2^k-c_1^k c_2^j) c_3^m c_4^n
\bar c_k \epsilon_{jmn} \nonumber \\
&\times&[(2uudd+2dduu-udud-uddu-duud-dudu)\bar s] \nonumber \\
&\times&[\{\uparrow\uparrow(\uparrow\downarrow+\downarrow\uparrow) -
(\uparrow\downarrow+\downarrow\uparrow)\uparrow\uparrow\}\downarrow
-(\uparrow\uparrow\downarrow\downarrow -
\downarrow\downarrow\uparrow\uparrow)\uparrow] \,\,\, ,
\label{eq:361}
\end{eqnarray}
\begin{eqnarray}
|({\bf 6},{\bf 6}, 0)({\bf
\bar 3},{\bf 6}, 1) \rangle  &=& \frac{1}{24 \sqrt{3}} 
(c_1^jc_2^k+c_1^k c_2^j) c_3^m c_4^n
\bar c_k \epsilon_{jmn} \nonumber \\ & \times &
[(2uudd+2dduu-udud-uddu-duud-dudu)\bar s] \nonumber \\
&\times&[(\uparrow\downarrow-\downarrow\uparrow)\uparrow\uparrow\downarrow
-\frac{1}{2}(\uparrow\downarrow-\downarrow\uparrow)(\uparrow\downarrow 
+ \downarrow\uparrow)\uparrow] \,\,\, ,
\label{eq:mixed1}
\end{eqnarray}
\begin{eqnarray}
|({\bf 6},{\bf \bar 3}, 1)({\bf \bar
  3},{\bf \bar 3}, 0) \rangle &=& \frac{1}{24} 
(c_1^j c_2^k+c_1^k c_2^j) c_3^m c_4^n 
\bar c_k \epsilon_{jmn} [(ud-du)(ud-du)\bar s] \nonumber \\
&\times& [\uparrow \uparrow (\uparrow\downarrow-\downarrow\uparrow)
\downarrow -\frac{1}{2}(\uparrow\downarrow+\downarrow\uparrow)
(\uparrow\downarrow-\downarrow\uparrow)\uparrow] \,\,\, .  
\label{eq:mixed2}
\end{eqnarray}
Here we have written the color wave function in tensor notation for
compactness, with $c^i \equiv (r,g,b)$.  The remaining component
states in Eq.~(\ref{eq:221}) can be obtained from
Eqs.~(\ref{eq:mixed1}) and (\ref{eq:mixed2}) by exchanging the first
and second pair of quarks. With these results, one may construct other
antidecuplet wave functions by application of SU(3) and isospin
raising and lowering operators.

It is often convenient for calculational purposes to have a decomposition
of the pentaquark wave function in terms of the quantum numbers of the
first three quarks, and of the remaining quark-antiquark pair. The
quark-antiquark pair can be either in a ${\bf 1}$ or ${\bf 8}$ of
color, which implies that we must have the same representations for
the three-quark ($q^3$) system, in order that a singlet may be formed.
As for flavor, the $q^3$ and $q\bar q$ systems must both be in ${\bf
8}$'s: the $q\bar q$ pair cannot be in a flavor singlet, since there
is no way to construct a ${\bf \overline{10}}$ from the remaining three
quarks, and the $q^3$ state must be an ${\bf 8}$ since the remaining
possibilities (${\bf 1}$ and ${\bf 10}$) do not yield an antidecuplet
when combined with the $q \bar q$ flavor octet.  Finally, the $q\bar
q$ spin can be either $0$ or $1$, which implies that the $q^3$ spin
can be either $1/2$ or $3/2$.  The states consistent with $q^3$
antisymmetry are then
\begin{center} \begin{tabular}{c}
$ |({\bf 1},{\bf 8}, 1/2)({\bf 1},{\bf 8}, 0) \rangle \,\,,\,\,
|({\bf 1},{\bf 8}, 1/2)({\bf 1},{\bf 8}, 1) \rangle \,\,,\,\,
|({\bf 8},{\bf 8}, 3/2)({\bf 8},{\bf 8}, 1) \rangle \,\,,\,\, $
\\
$ |({\bf 8},{\bf 8}, 1/2)({\bf 8},{\bf 8}, 0) \rangle \,\, ,\,\,
|({\bf 8},{\bf 8}, 1/2)({\bf 8},{\bf 8}, 1) \rangle $
\end{tabular} \end{center}
Again, we may find the coefficients by requiring that the total wave
function is antisymmetric under interchange of the four quarks.
Alternatively, we may take the overlap of any of these states with the
wave function that we have already derived in
Eqs.~(\ref{eq:221})-(\ref{eq:mixed2}).  The details and explicit
results will be presented in a longer publication~\cite{cckv2}. We
find
\begin{eqnarray}
| ({\bf 1},{\bf \overline{10}}, 1/2) \rangle &=&
\frac{1}{2}  |({\bf 1},{\bf 8}, 1/2)({\bf 1},{\bf 8}, 0) \rangle
+\frac{\sqrt{3}}{6} |({\bf 1},{\bf 8}, 1/2)({\bf 1},{\bf 8}, 1) \rangle
-\frac{\sqrt{3}}{3} |({\bf 8},{\bf 8}, 3/2)({\bf 8},{\bf 8}, 1) \rangle 
\nonumber \\
&+&\frac{1}{2} |({\bf 8},{\bf 8}, 1/2)({\bf 8},{\bf 8}, 0) \rangle
+\frac{\sqrt{3}}{6} |({\bf 8},{\bf 8}, 1/2)({\bf 8},{\bf 8}, 1) \rangle  
\,\,\, .
\label{eq:32}
\end{eqnarray}
Our sign conventions may be summarized by noting that each state on 
the right-hand side of Eq.~(\ref{eq:32}) contains the term 
$uudd\bar{s}\uparrow\uparrow\downarrow\uparrow\downarrow rbgr\bar{r}$
with positive coefficient.

Two interesting observations can be made at this point.  First,
Eqs.~(\ref{eq:221})-(\ref{eq:mixed2}) allow us to compute the
expectation value of $S_h=\sum_i |S_i|$, where $S_i$ is the
strangeness of the $i^{th}$ constituent.  This gives us the average
number of quarks in the state with either strangeness $+1$ or $-1$.
For the $Z^+$ state, the result is obviously $1$; Using the SU(3)
raising operator that changes $d\rightarrow s$ and ${\bar
s}\rightarrow -\bar d$, it is straightforward to evaluate the same
quantity for members of the antidecuplet with smaller total
strangeness.  We find
\begin{equation}
\langle Z^+| S_h |Z^+  \rangle = 3/3 \,\,,\,\,
\langle N_5  | S_h |N_5 \rangle = 4/3 \,\,,\,\, 
\langle \Sigma_5| S_h |\Sigma_5 \rangle = 5/3 \,\,,\,\, 
\langle \Xi_5| S_h |\Xi_5 \rangle = 6/3 \,\,\,,
\end{equation}
where $N_5$, $\Sigma_5$ and $\Xi_5$ represent the strangeness
$0$,$-1$ and $-2$ members of the ${\bf \overline{10}}$, respectively.
The nonstrange member of the  ${\bf \overline{10}}$ is heavier than
the $Z^+$ because it has, on average, $m_s/3$ more mass from
its constituent strange and antistrange quarks.

We also note that our decomposition in Eq.~(\ref{eq:32}) allows us to
easily compute overlaps with states composed of physical octet baryons
and mesons.  For example, the first term in Eq.~(\ref{eq:32}) may be
decomposed for the $Z^+$
\begin{equation}
|({\bf 1},{\bf 8}, 1/2)({\bf 1},{\bf 8}, 0) \rangle =
 \frac{1}{\sqrt{2}}(pK^0-nK^+) \,\,\,.
\end{equation}
The sizes of the coefficients of these terms affect the rate of the
``break-apart" decay modes, such as $Z^+ \rightarrow N K^+$.  We
therefore find that the smallness of the observed $Z^+$ decay width
($\alt 21$~MeV) does not originate with small group theoretic factors 
in the quark model wave function.

\section{Antidecuplet Masses and Decays}

Using the observed mass and width of the $Z^+$, one may make
predictions for the decay widths of other members of the antidecuplet.
Here we consider the decays ${\bf \overline{10}}\rightarrow B M$ where
$B$ ($M$) is a ground state octet baryon (meson).  We assume exact
SU(3)$_F$ symmetry in the decay amplitudes, but take into account
SU(3)$_F$ breaking in the mass spectra.  Mass splittings within the
antidecuplet obey an equal spacing rule when the strange quark mass is
the only source of SU(3)$_F$ breaking.  We compute these splittings
within the framework of the MIT bag model~\cite{bag1,bag2}, using the
original version for the sake of definiteness, including effects of
single gluon exchange interactions between the constituents.  (See
also \cite{CHP,CS}; these works show how the overall mass level of a
multiquark or gluonic state may be shifted, with only small changes in
the predictions for ground state baryons and for spin-dependent
splittings.) We also consider the possibility of dominant spin-isospin
constituent interactions, which would be expected if nonstrange
pseudoscalar meson exchange effects are important~\cite{glotz}.  The
predicted spectra differ significantly and yield distinguishable
patterns of kinematically accessible decays.

In the bag model, the mass of a hadronic state is given by
\begin{equation}
M = \frac{1}{R} \left\{ \sum \Omega_i - Z_0 + \alpha_s C_I \right\} 
+ B \frac{4 \pi R^3}{3}
\label{eq:bagmass}
\end{equation}
where $\Omega_i/R$ is the relativistic energy of the $i^{th}$ constituent 
in a bag of radius $R$, 
\begin{equation}
\Omega = (x^2 + m^2 R^2)^{1/2} \,\,\, ,
\end{equation}
and $x$ is a root of
\begin{equation}
\tan x = \frac{x}{1-m R - \Omega} \,\,\, .
\end{equation}
The parameter $Z_0$ is a zero-point energy correction, and $B$ is the
bag energy per unit volume.  In the conventional bag model, $Z_0 =
1.84$ and $B^{1/4}=0.145$~GeV.  The term $\alpha_s C_I$ represents the
possible interactions among the constituents. We first take into account the
color-spin interaction originating from single gluon exchange, so that
\begin{equation}
\alpha_s C_I = -\frac{\alpha_s}{4} \langle {\bf 1}, 
{\bf\overline{10}}, 1/2|  
\sum_{i<j} \mu(m_i,m_j) \, \lambda_i \cdot \lambda_j \, 
\sigma_i \cdot \sigma_j \, 
|{\bf 1}, {\bf\overline{10}}, 1/2 \rangle
\label{eq:scsc}
\end{equation}
where $\alpha_s=2.2$ is the value of the strong coupling appropriate
to the bag model, and $\mu(m_i,m_j)$ is a numerical coefficient that
depends on the masses of the of the $i^{th}$ and $j^{th}$ quarks.  For
the case of two massless quarks, $\mu(0,0) \approx 0.177$; the
analytic expression for arbitrary masses can be found in
Ref.~\cite{bag2}.

We take into account the effect of SU(3) breaking ({\em i.e.}, the
strange quark mass) in both $\Omega_i$ and in the coefficients
$\mu(m_i,m_j)$.  To simplify the analysis, we break the sum in
Eq.~(\ref{eq:scsc}) into two parts, quark-quark and quark-antiquark
terms, and adopt an averaged value for the parameter $\mu$ in each,
$\mu_{qq}$ and $\mu_{q\bar{q}}$.  Using the wave function in
Eqs.~(\ref{eq:221})-(\ref{eq:mixed2}) we find that the relevant
spin-flavor-color matrix elements are given by
\begin{eqnarray}
\langle {\bf 1}, {\bf\overline{10}}, 1/2|  
\sum_{i<j\neq 5} \lambda_i \cdot \lambda_j \, \sigma_i \cdot \sigma_j \, 
|{\bf 1}, {\bf\overline{10}}, 1/2 \rangle &=& 16/3 \nonumber \\
\langle {\bf 1}, {\bf\overline{10}}, 1/2| 
\sum_{i<j=5} \lambda_i \cdot \lambda_j \, \sigma_i \cdot \sigma_j \, 
|{\bf 1}, {\bf\overline{10}}, 1/2 \rangle &=& 40/3 \,\,\, ,
\end{eqnarray}
where $j=5$ corresponds to the antiquark.  This evaluation was done by
group theoretic techniques~\cite{cckv2}, as well as brute-force
symbolic manipulation~\cite{cgkm}.  To understand how we evaluate the
coefficients $\mu_{qq}$ and $\mu_{q\bar{q}}$ let us consider a
nucleon-like state in the antidecuplet, the $p_5$.  The probability of
finding an $s\bar{s}$ pair in the $p_5$ state is $2/3$. In this case,
$1/2$ of the possible $qq$ pairs will involve a strange quark. On the
other hand, the probability that the $p_5$ will contain five
non-strange constituents is $1/3$.  Thus, we take
\begin{equation}
\mu_{qq}(p_5) = \frac{2}{3}[\frac{1}{2}(\mu(0,0)+\mu(0,m_s))]
+\frac{1}{3} \mu(0,0) \,\,\, .
\end{equation}
By similar reasoning,
\begin{equation}
\mu_{q\bar{q}}(p_5)=\frac{1}{3}\mu(0,0)+\frac{1}{2}\mu(0,m_s)
+\frac{1}{6}\mu(m_s,m_s) \,\,\, .
\end{equation}
We also use the averaged kinetic energy terms
\begin{equation}
\frac{2}{3R}[3 \Omega(0)+2\Omega(m_s)] + \frac{1}{3R}[5 \Omega(0)] \,\,\, .
\end{equation}
The bag mass prediction is then obtained by numerically minimizing the
mass formula with respect to the bag radius $R$. Applying this
procedure to the $p_5$ and $Z^+$ states, we find the antidecuplet mass
splitting
\begin{equation}
\Delta M_{{\bf \overline{10}}} \approx 52 \mbox{ MeV}.
\end{equation}
We use the observed $Z^+$ mass, $1542$~MeV, and the splitting $\Delta
M_{{\bf \overline{10}}}$ to estimate the masses of the $p_5$,
$\Sigma_5$, and $\Xi_5$ states; we find $1594$, $1646$, and
$1698$~MeV, respectively.  Decay predictions from SU(3) symmetry are
summarized in Table~\ref{decaytable}.

While we used the bag model as a framework for evaluating the mass spectra  
above, we believe our results are typical of any constituent quark model.
 
We adopt a simpler approach in evaluating the effect of spin-isospin
constituent interactions,
\begin{equation}
\Delta M_{SI} = - C_\chi \langle {\bf 1}, {\bf\overline{10}}, 1/2|  
\sum_{i<j} \, \tau_i \cdot \tau_j \,\, \sigma_i \cdot \sigma_j \, 
|{\bf 1}, {\bf\overline{10}}, 1/2 \rangle  \,\,\, .
\label{eq:sisi}
\end{equation}
In this case the flavor generators $\tau$ are Pauli matrices, and the
coefficient $C_\chi = 25-30$~MeV is determined from the $N-\Delta$
mass splitting; we use $30$~MeV~\cite{glotz}.  The dimensionless matrix
element can be computed using Eqs.~(\ref{eq:221})-(\ref{eq:mixed2}),
and we find $10$, $20/3$, $25/9$ and $-5/3$ for the $Z^+$, $p_5$,
$\Sigma_5$ and the $\Xi_5$, respectively.  The mass splitting due to
the strange quark constituent mass can be estimated from our previous
bag model calculation, by excluding the spin-color interactions,
yielding $\Delta M_s \approx 55$~MeV.  Again fixing the $Z^+$ mass at
$1542$~MeV, we then find $1697$, $1869$, and $2058$~MeV for the $p_5$,
$\Sigma_5$, and $\Xi_5$ mass, respectively.  Decay results for this
mass spectrum are also presented in Table~\ref{decaytable}. Note that
a number of the decay modes that were kinematically forbidden before
are allowed if spin-isospin interactions dominate, due to the larger
predicted splitting within the antidecuplet. (For a smaller choice of
$C_\chi\approx 25$~MeV, the $\Sigma K$ modes are still inaccessible.)

\begin{table}[ht]
\begin{tabular}{lccc}
\hline\hline
Decay   \qquad\qquad & \qquad $|A/A_0|^2$ \qquad & 
\qquad $\Gamma/\Gamma_0$ (SC) \qquad\qquad &  $\Gamma/\Gamma_0$ (SI) \\
\hline 
$Z^+\rightarrow p K^0$ & $1$ &    $0.99$  & $0.99$ \\ 
$p_5\rightarrow \Lambda K^+$ & $1/2$ & -- & $0.49$ \\
$p_5\rightarrow p \eta$ &  $1/2$ & $0.50$& $0.68$ \\
$p_5\rightarrow \Sigma^+ K^0$ & $1/3$ & -- & $0.12$ \\
$p_5\rightarrow \Sigma^0 K^+$ & $1/6$ & -- &  $0.06$ \\
$p_5\rightarrow n \pi^+$ & $1/3$ & $0.63$ & $0.68$  \\
$p_5\rightarrow p \pi^0$ & $1/6$ & $0.32$ & $0.34$ \\
$\Sigma_5^+ \rightarrow \Xi^0 K^+$ & $1/3$ & -- &  $0.30$ \\
$\Sigma_5^+ \rightarrow \Sigma^+ \eta$ & $1/2$  & -- & $0.62$ \\
$\Sigma_5^+ \rightarrow \Lambda \pi^+$ & $1/2$ & $0.89$ & $1.11$ \\
$\Sigma_5^+ \rightarrow  p \bar K^0$ & $1/3$ & $0.45$ & $0.63$ \\
$\Sigma_5^+ \rightarrow \Sigma^+ \pi^0$ & $1/6$ & $0.27$ & $0.36$\\
$\Sigma_5^+ \rightarrow \Sigma^0 \pi^+$ & $1/6$ & $0.27$ & $0.36$ \\
$\Xi_5^+ \rightarrow \Xi^0 \pi^+$ & $1$ & $1.47$ & $2.37$ \\
$\Xi_5^+ \rightarrow \Sigma^+ \bar K^0$ & $1$ & $0.36$ & $1.99$ \\
\hline\hline 
\end{tabular}
\caption{SU(3) decay predictions for the highest isospin members of the 
antidecuplet. $A_0$ and $\Gamma_0$ are the amplitude and partial
decay width for $Z^+ \rightarrow N K^+$, respectively; SC and SI indicate 
antidecuplet mass spectra assuming dominant spin-color or spin-isospin 
constituent interactions.}
\label{decaytable}
\end{table}      

The Skyrme model also has predictions~\cite{diakonov} for the masses 
and decays of the antidecuplet.  The mass splittings there were about 
180 MeV between each level of the decuplet (with the $Z^+$ still the 
lightest), considerably larger splittings than we find in a constituent 
quark model where the mass splittings come from strange quark masses 
and from color-spin interactions.  Mass splittings using isospin-spin 
interactions were, on the other hand, more comparable to the Skyrme 
model results.

Decays of the antidecuplet into a ground state octet baryon and an 
octet meson involve a decay matrix element and phase space.  Ratios of 
decay matrix elements for pure antidecuplets, such as we show in Table 
I, are fixed by $SU(3)_F$ symmetry.  They are the same in any model, as 
may be confirmed by comparing Table I to results in~\cite{diakonov}.  
We have neglected mixing; Ref.~\cite{diakonov} does consider mixing but 
does not find large consequences for the decays.  The differences 
between relative decay predictions are then due to differences in phase 
space, and the differences are due to masses and due to parity.  
Negative parity states decaying to ground state baryon and pseudoscalar 
meson have S-wave phase space, while positive parity states have P-wave 
phase space.  Note also that $SU(3)_F$ symmetry does not allow decays 
of antidecuplets into decuplet baryons plus octet mesons.
\section{Discussion}

In this letter we have shown how to construct the quark model wave
functions for members of the pentaquark antidecuplet, the flavor
multiplet that we argue is most likely to contain the strangeness one
state recently observed in a number of
experiments~\cite{nakano,barmin,stepanyan}.  We present two
decompositions of the ${\bf \overline{10}}$ wave function that are
useful for computing spin-flavor-color matrix elements, and that
reveal the hidden strangeness in each component state.  In addition,
we have presented the $Z^+$ wave function in explicit form.  We use
these results to estimate the effect of spin-color and spin-isospin
interactions on the pentaquark mass spectrum.  In the first case, we
use the MIT bag as a representative constituent quark model to compute
the equal spacing between antidecuplet states that differ by one unit
of strangeness; we estimate a splitting of $52$~MeV.  The observed
$Z^+$ mass and SU(3) symmetry then allows us to make decay
predictions.  Notably, if only color-spin interactions are present,
decays of the $p_5$ and $\Sigma_5$ to final states in which both decay
products have nonzero strangeness are kinematically forbidden. In 
addition, the $\Xi_5$ states are narrower than those in Ref.~\cite{diakonov}, 
so that experimental detection might be possible and dramatic. If instead, 
spin-isospin interactions dominate, all the decays in Table~\ref{decaytable} 
become kinematically accessible. 

The work summarized here sets the groundwork for further
investigation.  Of particular interest to us is the relation between
bag model predictions for the absolute pentaquark mass (rather than
the mass splittings considered here) and the mass of other multiquark
exotic states.  The conventional MIT bag predicts a $Z^+$ mass that is
too large relative to the experimental value (we find that a
prediction of about $1700$~MeV is typical); however, these numbers can
be easily reconciled by allowing bag model parameters 
to float~\cite{CHP,CS}. An appropriate analysis requires a
simultaneous fit to pentaquark and low-lying non-exotic hadron masses,
and consideration of center-of-mass corrections.  Whether such fits
simultaneously allow for sufficiently heavy six-quark states, given a
choice of constituent interactions, is an open question.  Our analysis also 
gives insight into other pentaquark states.  For example,
there are nucleon-like states in the pentaquark octet (states in the 
same spin-color representation as the $Z^+$) which are potentially light.
However, we find that these states also have hidden strangeness, placing them 
within one-third of the strange quark mass below the $Z^+$, if no other 
effects are considered, and at or above the $Z^+$ mass if spin-isospin
interactions are taken into account. This is one example of the value of 
extending our present analysis to other pentaquark multiplets.  A more 
detailed discussion of these topics, as well as of the group theoretical 
issues described here will be presented in a longer publication~\cite{cckv2}.




\begin{acknowledgments}
We thank Robert Jaffe for useful comments.  We thank the NSF for
support under Grant Nos.\ PHY-0140012, PHY-0243768 and PHY-0245056.
\end{acknowledgments}


\end{document}